\documentclass[conference]{IEEEtran}
\IEEEoverridecommandlockouts

\usepackage{pifont}
\usepackage{cite}
\usepackage{amsmath,amssymb,amsfonts}
\usepackage{algorithmic}
\usepackage{graphicx}
\usepackage{textcomp}
\usepackage{xspace}
\usepackage{xcolor}
\usepackage{url}
\usepackage{capt-of}
\usepackage[most]{tcolorbox}
\definecolor{BetterTeal}{RGB}{46,92,184}
\definecolor{WorseOrange}{RGB}{176,112,58}
\newcommand{\goodup}[1]{\textcolor{BetterTeal}{$\uparrow$\,#1}}
\newcommand{\gooddown}[1]{\textcolor{BetterTeal}{$\downarrow$\,#1}}
\newcommand{\badup}[1]{\textcolor{WorseOrange}{$\uparrow$\,#1}}
\newcommand{\baddown}[1]{\textcolor{WorseOrange}{$\downarrow$\,#1}}
\newcommand{\cellchange}[2]{\begin{tabular}[c]{@{}c@{}}#1\\[-1pt]{\tiny #2}\end{tabular}}
\newcommand{\sigone}{\textsuperscript{*}}
\newcommand{\sigtwo}{\textsuperscript{**}}
\newtcolorbox{findingbox}[1]{
  enhanced,
  breakable,
  colback=blue!4!white,
  colframe=blue!65!black,
  colbacktitle=blue!80!black,
  coltitle=white,
  fonttitle=\bfseries,
  fontupper=\normalsize,
  title={#1},
  boxrule=0.9pt,
  arc=2pt,
  left=6pt,
  right=6pt,
  top=6pt,
  bottom=6pt,
  before skip=6pt,
  after skip=8pt
}
\newcommand{\approach}{{SchrodingerRepo}\xspace}
\def\BibTeX{{\rm B\kern-.05em{\sc i\kern-.025em b}\kern-.08em
    T\kern-.1667em\lower.7ex\hbox{E}\kern-.125emX}}

\begin{document}

\title{Schr\"odinger's Code Repository: Have LLMs Learned SWE-bench or Memorized It?}

\author{
\IEEEauthorblockN{Silin Chen\textsuperscript{1,*}, Yufei Yang\textsuperscript{2,*}, Xiaodong Gu\textsuperscript{1,\textdagger}, Yuling Shi\textsuperscript{1}, Chengcheng Wan\textsuperscript{3}, Haibing Guan\textsuperscript{1}}
\IEEEauthorblockA{\textsuperscript{1}Shanghai Jiao Tong University\\
\texttt{cslsolow@gmail.com, \{xiaodong.gu,yuling.shi,hbguan\}@sjtu.edu.cn}}
\IEEEauthorblockA{\textsuperscript{2}Xi'an Jiaotong University\\
\texttt{qfrfyflc@stu.xjtu.edu.cn}}
\IEEEauthorblockA{\textsuperscript{3}East China Normal University, Shanghai Innovation Institute\\
\texttt{ccwan@sei.ecnu.edu.cn}}
\thanks{\textsuperscript{*}Silin Chen and Yufei Yang contributed equally to this work.}
\thanks{\textsuperscript{\textdagger}Xiaodong Gu is the corresponding author.}
}

\maketitle
\begin{abstract}
Repository-level coding benchmarks have become the primary standard for evaluating coding agents. However, these benchmarks inherently suffer from data leakage because they are built upon popular open-source repositories that are repeatedly used for training. A static repository representation makes it difficult to determine whether strong performance reflects robust repository reasoning or memorization of canonical repository cues.
To address this limitation, we propose \approach (Schrödinger's Repository), a novel evaluation framework that rigorously tests the true comprehension of coding agents. Instead of repeatedly using a static representation of the test repository, \approach treats the test repository as an evaluation-time latent variable that is dynamically instantiated only when the agent enters the evaluation environment. This approach yields a semantically equivalent repository that preserves the original executable behavior, while eroding familiar repository-side cues such as naming conventions, file layouts, or idiosyncratic implementation patterns. Specifically, \approach comprises four transformation levels: problem statement reconstruction, namespace remapping, intra-file layout reordering, and functionality-preserving code rewriting.
We evaluate popular LLMs on SWE-bench Verified and SWE-QA with \approach. Our experiments reveal several key findings. First, removing familiar repository cues consistently degrades agent performance while significantly increasing interaction costs across all evaluated LLMs. Further analyses show that this additional cost is primarily driven by the agents' newly exposed struggle with repository exploration. These results suggest that the strong performance of current agents partially reflects the memorization of surface-level repository cues, highlighting the importance of evaluating agents under dynamically instantiated repository representations\footnote{Our code and data are available at \url{https://github.com/cslsolow/Schrodinger-Repo}}.
\end{abstract}



\section{Introduction}
Evaluating coding agents~\cite{wang2025openhands, antoniades2025swe,zeng2025pruning,zeng2026dockerless,zeng2026glimprouter,hu2026line,gao2026swe,chang2026test,202605.2065,gu2018deep,chen2026repo0designdrivenzerotoallcode,chen2026skillforgeselfdistillingagentsprojectspecific,lin2026knowfixqadrivenrepository,huang2027planning,chen2026rethinkingvalueagentgeneratedtests,ma2026llmagentscoderepositories,lin2024llmscontinuouslearnersimproving} has become increasingly challenging as software engineering tasks require models to reason over large codebases, navigate project structure and documentation, localize root causes, coordinate edits across multiple files, and validate fixes in executable environments. These properties make repository-level evaluation a particularly demanding and practically meaningful setting for assessing modern coding agents. 
As a representative benchmark for repository-level issue resolution, SWE-bench~\cite{jimenez2024swe} has become the standard for evaluation in this domain. It curates 2,294 real-world GitHub issues, complete with executable environments and test-based evaluation. Furthermore, SWE-bench Verified~\cite{jimenez2024swe} provides a manually curated subset of 500 instances, which is now widely adopted for evaluating frontier agents.

\begin{figure}[t]
  \centering
  \includegraphics[width=\linewidth]{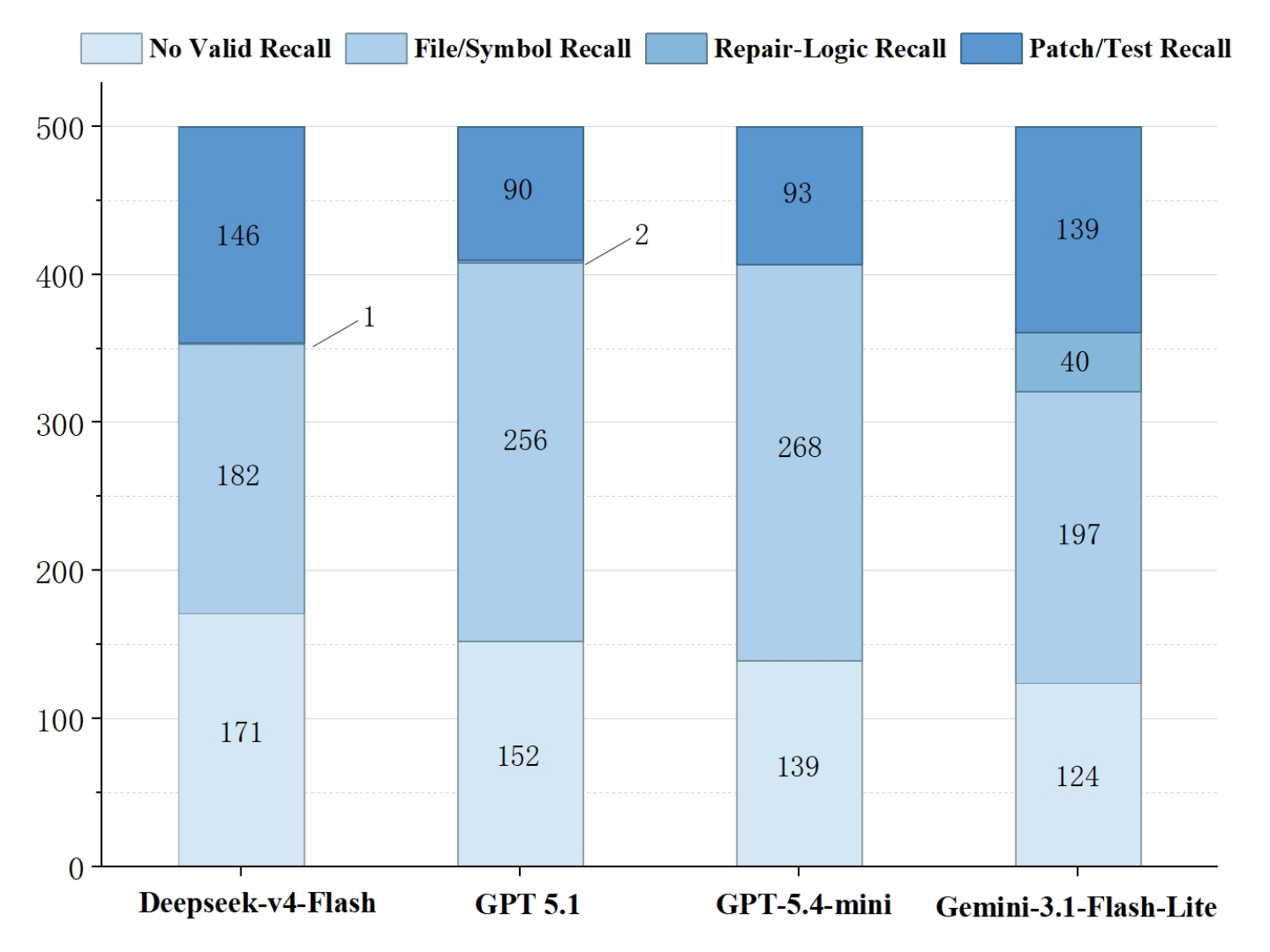}
  \caption{Motivation experiment: human-judged evidence of task-specific memorization on SWE-bench Verified.}
  \label{fig:background}
  \vspace{-8pt}
\end{figure}

Despite its widespread adoption, existing SWE benchmarks have been known to suffer from data leakage \cite{openai2026retireverified}. SWE-bench is built upon widely used open-source repositories. Each issue is exhibited through a single canonical repository presentation that is repeatedly encountered across training, development, and evaluation. Consequently, high benchmark scores may merely reflect a model's memorization of repository-specific patterns (e.g., naming conventions and APIs) rather than true code reasoning. This ambiguity leaves a fundamental question unanswered: \textit{does strong SWE-bench performance reflect robust software engineering capabilities, or simply familiarity with leaked repository representations?}

Recent work has attempted to mitigate these concerns by constructing continuously updated benchmarks, including SWE-bench Live~\cite{zhang2026swe}, SWE-rebench~\cite{badertdinov2026swe}, and SWE-bench Pro~\cite{deng2025swe}. These benchmarks continuously incorporate newly released instances and focus on issues created after the release of up-to-date models to reduce direct contamination. However, they remain inherently limited in scale and issue-type coverage, making it difficult to fully capture the diversity of real-world repository-level issues.


To address this limitation, we propose \approach (Schr\"odinger's Repository), a novel evaluation framework that rigorously tests the robustness of coding agents. 
Instead of repeatedly using a static representation of the test repository, \approach introduces controlled transformations to existing repositories, yielding a semantically equivalent codebase that preserves the original executable behavior while eroding familiar repository-side cues such as naming conventions, file layouts, and idiosyncratic implementation patterns. By varying a random seed, our framework renders a deterministic yet distinct view of the repository for every evaluation run. In this sense, repository representation is no longer a fixed benchmark artifact but a latent state whose concrete realization is determined only when an agent enters the evaluation environment. 

Concretely, \approach systematically transforms an evaluation instance through four transformation levels: Level~1 reconstructs the problem statement, Level~2 remaps repository-owned namespaces, Level~3 reorders intra-file layout, and Level~4 rewrites local implementations while preserving functionality.

We evaluate GPT-5.4-mini, GPT 5.1, DeepSeek-v4-Flash, and Gemini-3.1-Flash-Lite under controlled repository transformations generated by \approach
on SWE-bench Verified~\cite{jimenez2024swe}, the March 2026 SWE-rebench Leaderboard~\cite{badertdinov2026swe} split containing instances created after the release of the evaluated LLMs, and SWE-QA~\cite{peng2026swe}. Our study yields the following findings:

%
%
%

\ding{182} Current coding agents exhibit substantial dependence on repository-side cues. On SWE-bench Verified, the full transformed setting reduces Pass@1 by 6.0--14.4 percentage points across the evaluated models, and among individual transformation levels, Namespace Mapping produces the largest drop.

\ding{183} Removing familiar repository-side cues forces coding agents to spend substantially more interaction budget on repository exploration and localization. Across the evaluated agents, 81.6--83.6\% of the additional actions are spent on exploration-oriented behaviors, accompanied by markedly higher token consumption.

\ding{184} The effects of repository representation are not limited to issue resolution, but generalize to broader repository-level tasks. On SWE-QA, transformed repository views reduce answer quality by up to 4.64 points while increasing actions by 18.15--43.02\%.

\ding{185} On temporally held-out SWE-rebench instances, repository transformations preserve Pass@1 while still increasing interaction cost. This suggests that the observed degradation on SWE-bench Verified is not simply caused by making tasks intrinsically harder, but by removing familiar repository-side cues that current agents rely on.

Overall, this paper presents the first systematic study of repository representation sensitivity in repository-level coding-agent evaluation. 
This design enables controlled measurement of whether benchmark performance reflects robust repository-level reasoning or reliance on familiar canonical repository cues, contributing to more reliable and interpretable evaluation of repository-level software engineering agents.

\section{Background}

SWE-bench Verified~\cite{jimenez2024swe} has become a central benchmark for repository-level coding agents, but its instances are drawn from public, widely used repositories whose issues, code, tests, and discussions may appear in model training data. Following OpenAI's analysis of why SWE-bench Verified no longer reliably measures frontier coding capabilities~\cite{openai2026retireverified}, we first conduct a motivation experiment to measure whether evaluated models exhibit task-specific memory before interacting with the repository.

In this experiment, human experts decompose each instance's problem statement from the complete issue description into semantic units ordered from broad to specific, and reveal these units to the evaluated LLM round by round. At the beginning, the evaluated LLM only observes the instance ID and a small number of issue-level semantic units; it cannot access repository files, the gold patch, or test information. After each round, human experts compare the evaluated LLM's output against hidden reference information and determine whether it contains task-specific content that has not appeared in the current prompt. The resulting evidence is grouped into four categories: no valid recall indicates no effective task-specific recall, file/symbol recall indicates recovery of affected files or symbols, repair-logic recall indicates recovery of the core fix logic, and patch/test recall indicates recovery of concrete patch content, modified code lines, or test-specific information. Based on this judgment, the human experts decide whether to continue revealing additional semantic units or stop and request more concrete recall evidence. Figure~\ref{fig:background} summarizes the resulting leakage evidence over SWE-bench Verified. For each evaluated model, more than 65\% of instances exhibit clear data-leakage evidence, and more than 18\% of instances can be recalled at the patch/test level.

    

\section{Approach}

\subsection{Overview} 
\approach transforms each SWE-bench Verified task from a fixed canonical repository presentation into an evaluation-time repository view that is semantically equivalent but not observable before the agent enters the environment. Figure~\ref{fig:method-overview} shows the overall design. The goal is to erode repository-side cues that may have been memorized from public benchmark artifacts, while preserving the underlying issue, executable behavior, and test-defined correctness criteria. To this end, \approach builds a seeded and invertible mapping between the original repository and an agent-facing view, then applies semantics-preserving transformations at both the observation layer and the working-repository layer. 

Level 1 reconstructs the problem statement to reduce dependence on canonical wording. Level 2 remaps repository identity, paths, and repository-owned symbols to alter familiar namespace cues. Levels 3 and 4 operate on the repository state by producing structurally or behaviorally equivalent code variants that still satisfy the original execution and testing constraints. Because the mapping is invertible, tool execution remains grounded in the real SWE-bench environment and final patches can be translated back into the original repository coordinates for standard evaluation.

The four levels can be evaluated either jointly, to test end-to-end robustness under combined representation changes, or individually, to isolate the effect of each transformation type. The following four subsections describe these components in turn: 

\begin{figure*}[t]
    \centering
    \includegraphics[width=0.90\textwidth]{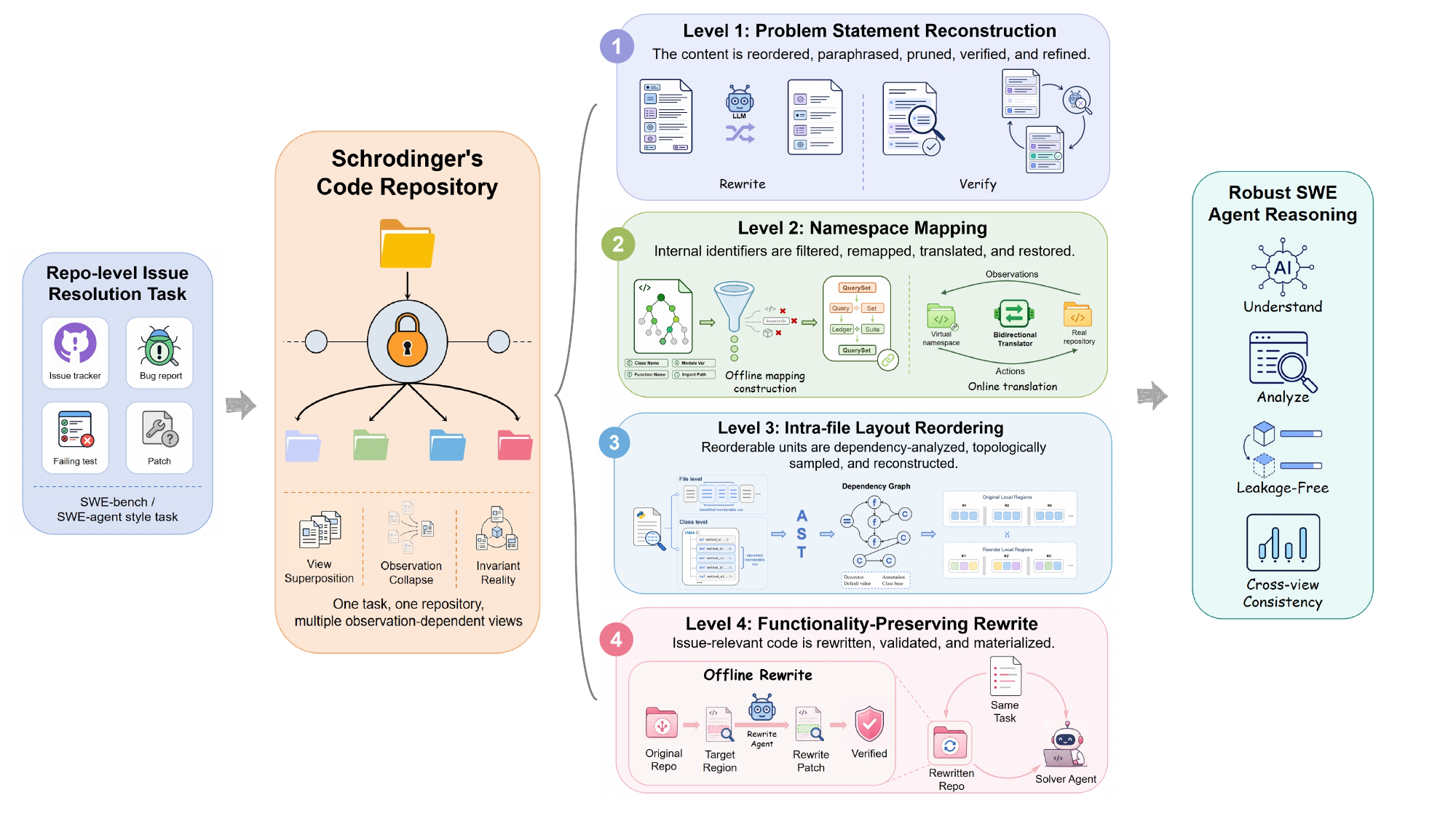}
    \caption{Overview of \approach.}
    \label{fig:method-overview}
\end{figure*}


\subsection{Level 1: Problem Statement Reconstruction}
To reduce benchmark-specific lexical cues in the natural-language task description without altering the underlying bug-fixing objective, Level 1 reconstructs the original problem statement into a semantically equivalent variant. The goal is to reduce sensitivity to canonical benchmark phrasing that may arise from repeated exposure to static instances, while preserving the specification-level semantics of the task, including the bug description, functional requirements, and success criteria.

Level 1 applies a two-stage LLM-based transformation over the problem statement. A generator first produces a rewritten version by reordering information, paraphrasing expressions, and removing non-essential details that do not affect task semantics, such as identifiers or incidental metadata when not functionally required. A verifier LLM then checks whether the reconstructed statement preserves all task-defining constraints; if information is missing or weakened, it triggers a refinement step to restore the omitted semantics.



\subsection{Level 2: Namespace Mapping}
Whereas Level 1 transforms only the natural-language problem statement, Level 2 targets repository-side lexical cues while preserving the underlying executable task. To mitigate repository-specific namespace leakage without altering functional behavior, Level 2 introduces a seeded and invertible virtual namespace over repository paths, modules, and symbols. From the agent’s perspective, the repository is fully renormalized into this virtual namespace, while the execution backend continues to operate on the original SWE-bench Verified instance. All transformations occur at the observation and interaction level without modifying the physical repository.

Operationally, Level 2 consists of an offline mapping construction stage and an online translation stage. Offline, \approach constructs a repository-level mapping bundle by first extracting candidate identifiers from the abstract syntax tree (AST)~\cite{noonan1985algorithm} of the codebase. All symbol-bearing AST nodes—including class definitions, function definitions, module-level variables, and import references—are traversed to collect repository-relevant identifiers. To ensure executability and prevent semantic drift beyond repository boundaries, we apply a strict filtering procedure that removes Python built-in functions, reserved keywords, and third-party library symbols, including external API roots and imported package namespaces. The resulting identifier set is therefore restricted to repository-internal terms that encode domain-specific semantics.

The filtered identifiers are then decomposed into subword-level tokens and mapped to semantically plausible alternatives under a seed-controlled procedure. During reconstruction, the system preserves common naming conventions including CamelCase class names, snake\_case function or file names, UPPER\_CASE constants, and dotted module-path structure. This mapping also enforces consistency at the token level, such that shared subcomponents across multiple identifiers are translated coherently. For example, \texttt{QuerySet} may be decomposed into \texttt{Query}+\texttt{Set} and remapped as \texttt{Ledger}+\texttt{Suite}, yielding \texttt{LedgerSuite}; the same \texttt{Query}$\rightarrow$\texttt{Ledger} mapping can then be reused in other identifiers that contain the token \texttt{Query}. The resulting bundle defines a structured, repository-specific lexicon that induces a consistent virtual namespace over all internal symbols, file paths, and module references, while preserving external APIs and language-level constructs unchanged. Multiple mapping variants may be generated per repository, with deterministic selection based on repository identity and semantic seed.

During execution, the mapping is instantiated as a bidirectional translator between the real repository and the agent-visible environment. Observations (e.g., code context, problem statements, and execution traces) are translated into the virtual namespace, while agent actions are translated back into the original namespace before execution. This guarantees that all execution remains grounded in the original repository state, while the agent operates entirely within the transformed representation space.

\subsection{Level 3: Intra-file Layout Reordering}
Unlike Levels 1 and 2, which operate on the agent-visible observation space, Level 3 modifies the underlying repository state prior to task execution. It constructs a semantics-preserving variant of the repository in which only intra-file ordering is transformed, while all functional behavior remains unchanged. This allows us to isolate whether agents rely on canonical code ordering as an implicit structural prior in repository-level reasoning.

Concretely, Level 3 applies reordering within contiguous runs of reorderable definitions at both the file and class levels, including top-level function and class definitions as well as method definitions within class bodies. Non-reorderable statements act as structural anchors that partition reorderable regions, ensuring that only local ordering is affected while higher-level organization is preserved.

For each reorderable run, Level 3 analyzes the corresponding abstract syntax tree and constructs a definition-time dependency graph $G$ over reorderable units~\cite{noonan1985algorithm}. Each unit is modeled as a node, and directed edges encode definition-time name availability constraints induced by constructs such as decorators, default argument values, type annotations, class bases, and class-body expressions. A directed edge is introduced whenever one unit depends on names defined by another unit within the same run, ensuring that any valid ordering preserves import-time and class-construction semantics.

Given $G$, Level 3 samples a valid ordering via randomized topological sorting over the induced partial order. If multiple valid orderings exist, one is selected uniformly at random; if no alternative ordering exists, the original sequence is retained. The reordered run is then rendered back into source form, while all non-reorderable regions remain unchanged.

The resulting repository is materialized as an execution-time overlay used for agent interaction, while evaluation is always grounded in the original SWE-bench repository state.

\subsection{Level 4: Functionality-Preserving Rewrite}
Whereas Level 3 changes only the relative ordering of existing definitions, Level 4 changes local implementation form itself. Its goal is to expose whether agents rely on memorized implementation patterns near the fix location rather than reasoning over task semantics. To this end, Level 4 rewrites issue-relevant code into behaviorally equivalent variants before the downstream issue-resolution agent begins solving the task. The rewritten repository is then used as the working environment for standard SWE-bench Verified evaluation.

Level 4 follows a two-stage workflow. In the first stage, the system constructs candidate rewrites offline. For each instance, it first identifies the code region most directly tied to the original repair signal, then invokes a constrained rewriting agent to produce a unified diff patch whose purpose is not to solve the issue, but to restate the existing implementation in a behaviorally equivalent yet substantially different form. The rewriting objective is therefore representation change rather than bug fixing: the transformed code should preserve functionality while altering the local implementation patterns that an agent would otherwise observe near the eventual fix location.

In the second stage, validated rewrites are materialized into the working repository and presented to the downstream issue-resolution agent. As a result, the agent no longer interacts with the canonical implementation form of the original environment, but with a rewritten variant that preserves the same unresolved task. This design makes Level 4 complementary to Level 3. Level 3 changes the order in which definitions are encountered, whereas Level 4 changes the implementation form of issue-relevant code itself. Together, they test whether agent performance is robust not only to changes in repository organization, but also to changes in the local coding patterns surrounding the bug.

\subsection{Level-wise Validity Checks}
\label{subsec:level-wise-validity}
We validate each level separately so that benchmark outcomes reflect the downstream agent's issue-resolution ability rather than artifacts introduced by the transformation itself. 
For each transformation level, human reviewers additionally inspected 100 randomly sampled instances and confirmed that the transformed and original versions refer to the same underlying task, preserve the target issue, and expose equivalent information needed for issue resolution.
For Level 1, a human engineer performs a final review to ensure that the reconstructed statement is semantically equivalent to the original, before it is presented to the downstream agent.
For Level 2, validity is enforced at the translation interface rather than by rewriting the repository. The translator preserves the command head and rewrites only namespace-bearing arguments or embedded code payloads, leaving bash-command semantics unchanged. A session notebook records the virtual-to-real substitutions actually instantiated during the run, and reverse translation is restricted to these observed mappings. Together, these safeguards change the agent-visible namespace without changing the executable task.

For Levels 3 and 4, validity is checked at the repository level in the official SWE-bench execution environment by running the corresponding test suite. Let $V$ denote a candidate transformed repository produced either by Level 3 reordering or by a Level 4 rewrite. We retain $V$ only if it satisfies $\mathrm{Pass\_to\_Pass}(V)=1$ and $\mathrm{Fail\_to\_Pass}(V)=0$, where the first condition preserves already-correct behavior and the second ensures that the target bug remains unfixed. Together, these constraints ensure that the transformed repository remains the same unresolved SWE-bench Verified instance.


\subsection{Patch Recovery}

We recover the final submission from repository states rather than from raw agent-emitted patch text. Let $B$ denote the original repository state for a benchmark instance, and let $R'$ denote the final repository state after all enabled transformations and agent edits. \approach reconstructs the submission as
\[
\mathrm{final\_submission}
= \bigoplus_{f \in C_B} \mathrm{Diff}(b_f, r'_f),
\]
where $C_B$ is the set of files whose final contents differ from the original repository, $b_f$ is the original contents of file $f$, and $r'_f$ is its final contents. This formulation ensures that evaluation is consistently grounded in the original SWE-bench repository state, regardless of whether the agent operates on a transformed or rewritten representation. It unifies evaluation across all transformation levels by deriving submissions directly from final repository states.

\section{Repository Representation Study}

Our research is guided by four research questions: 

\noindent\textbf{RQ1: To what extent do repository-side representations influence coding agent performance on repository-level issue resolution? }
We compare agent performance on canonical issue-resolution instances with transformed variants in which all four levels are enabled jointly, and we additionally evaluate each level separately to identify which representation changes contribute most to the robustness gap.

\noindent\textbf{RQ2: How does eroding familiar repository-side surface cues affect coding agent on repository-level issue resolution?}
We analyze how transformed repository views change agent behavior by redistributing actions across a fine-grained action taxonomy: \texttt{navigate}, \texttt{search}, \texttt{read}, \texttt{probe}, \texttt{edit}, and \texttt{test}.

\noindent\textbf{RQ3: Can \approach transfer to other repository-level tasks?}
We apply the same transformation framework to repository-level question answering and evaluate whether similar performance and behavioral shifts appear beyond issue resolution.

\noindent\textbf{RQ4: Does \approach make issue-resolution instances intrinsically more difficult?}
We evaluate only the full \approach setting on temporally held-out issue-resolution instances created after the release of the evaluated LLMs, to examine whether the same effect persists when direct exposure to the evaluated instances is unlikely.

Answering these questions requires a controlled way to vary repository presentation while holding the underlying task, execution semantics, and evaluation criteria fixed. We therefore construct \approach as the experimental instrument for this study. It enables us to systematically manipulate the agent-facing representation of a benchmark instance without changing the underlying repository behavior, allowing us to isolate how repository-side cues affect agent effectiveness, efficiency, and behavior.

\begin{table*}[!t]
\centering
\caption{Results on SWE-Bench Verified.}
\label{tab:rq1-main}
\renewcommand{\arraystretch}{1.1}
{\scriptsize
\setlength{\tabcolsep}{4pt}
\begin{tabular*}{0.92\textwidth}{@{\extracolsep{\fill}}l c c c c}
\hline
Setting & Pass@1 & Avg. Actions & Avg. Input tokens & Avg. Output tokens \\
\hline
\multicolumn{5}{c}{\textbf{GPT 5.1}} \\
\hline
Baseline & 44.6\% & 20.48 & 178,247 & 2,181 \\
Level 1 & 44.6\% (\gooddown{0.00\%}) & 20.31 (\gooddown{0.83\%}) & 177,102 (\gooddown{0.64\%}) & 2,169 (\gooddown{0.55\%}) \\
Level 2 & 37.2\%\sigtwo (\gooddown{7.4\%}) & 33.45\sigtwo (\goodup{63.3\%}) & 467,563\sigtwo (\goodup{162.3\%}) & 3,621\sigtwo (\goodup{66.0\%}) \\
Level 3 & 43.0\% (\gooddown{1.6\%}) & 21.81 (\goodup{6.5\%}) & 203,968 (\goodup{14.4\%}) & 2,290 (\goodup{5.0\%}) \\
Level 4 & 43.4\% (\gooddown{1.2\%}) & 21.68 (\goodup{5.8\%}) & 203,535 (\goodup{14.2\%}) & 2,492\sigtwo (\goodup{14.2\%}) \\
\textbf{\approach} & \textbf{36.2\%}\sigtwo (\gooddown{8.4\%}) & \textbf{33.09}\sigtwo (\goodup{61.5\%}) & \textbf{465,712}\sigtwo (\goodup{161.3\%}) & \textbf{3,678}\sigtwo (\goodup{68.6\%}) \\
\hline
\multicolumn{5}{c}{\textbf{GPT-5.4-mini}} \\
\hline
Baseline & 46.8\% & 11.25 & 65,905 & 1,564 \\
Level 1 & 46.8\% (\gooddown{0.0\%}) & 11.56 (\goodup{12.78\%}) & 66,973 (\goodup{1.62\%}) & 1,563 (\baddown{0.03\%}) \\
Level 2 & 40.4\%\sigtwo (\gooddown{6.4\%}) & 14.92\sigtwo (\goodup{32.6\%}) & 137,190\sigtwo (\goodup{108.2\%}) & 2,128\sigtwo (\goodup{36.1\%}) \\
Level 3 & 43.4\%\sigone (\gooddown{3.4\%}) & 13.27\sigtwo (\goodup{17.9\%}) & 106,676\sigtwo (\goodup{61.9\%}) & 1,979\sigtwo (\goodup{26.6\%}) \\
Level 4 & 44.6\%\sigone (\gooddown{2.2\%}) & 12.66\sigtwo (\goodup{12.5\%}) & 92,138\sigtwo (\goodup{39.8\%}) & 1,961\sigtwo (\goodup{25.3\%}) \\
\textbf{\approach} & \textbf{35.6\%}\sigtwo (\gooddown{11.2\%}) & \textbf{19.83}\sigtwo (\goodup{76.2\%}) & \textbf{234,216}\sigtwo (\goodup{255.4\%}) & \textbf{2,915}\sigtwo (\goodup{86.4\%}) \\
\hline
\multicolumn{5}{c}{\textbf{DeepSeek-v4-Flash}} \\
\hline
Baseline & 72.8\% & 46.19 & 1,046,985 & 14,505 \\
Level 1 & 70.8\% (\gooddown{2.0\%}) & 61.54\sigtwo (\goodup{33.2\%}) & 1,494,954\sigtwo (\goodup{42.8\%}) & 13,583\sigtwo (\baddown{6.4\%}) \\
Level 2 & 66.8\%\sigtwo (\gooddown{6.0\%}) & 98.13\sigtwo (\goodup{112.4\%}) & 3,332,478\sigtwo (\goodup{218.3\%}) & 22,549\sigtwo (\goodup{55.5\%}) \\
Level 3 & 72.0\% (\gooddown{0.8\%}) & 47.52 (\goodup{2.9\%}) & 1,099,395 (\goodup{5.0\%}) & 14,861 (\goodup{2.5\%}) \\
Level 4 & 70.0\%\sigone (\gooddown{2.8\%}) & 48.88 (\goodup{5.8\%}) & 1,227,483 (\goodup{17.2\%}) & 16,848\sigone (\goodup{16.2\%}) \\
\textbf{\approach} & \textbf{66.8\%}\sigtwo (\gooddown{6.0\%}) & \textbf{99.78}\sigtwo (\goodup{116.0\%}) & \textbf{3,702,241}\sigtwo (\goodup{253.6\%}) & \textbf{26,347}\sigtwo (\goodup{81.6\%}) \\
\hline
\multicolumn{5}{c}{\textbf{Gemini-3.1-Flash-Lite}} \\
\hline
Baseline & 56.7\% & 51.01 & 1,141,585.57 & 10,114.90 \\
\textbf{\approach} & \textbf{42.3\%} (\gooddown{14.4\%}) & \textbf{92.37} (\goodup{81.09\%}) & \textbf{2,694,826.90} (\goodup{136.06\%}) & \textbf{16,958.64} (\goodup{67.66\%}) \\
\hline
\multicolumn{5}{r}{\sigone: $p-value<0.05$, \sigtwo: $p-value<0.01$.} \\
\end{tabular*}
}
\end{table*}

%
\subsection{LLM Selection}
We instantiate the representative agent with mini-swe-agent~\cite{yang2024swe}, which has been widely used as a scaffold by recent coding-agent methods~\cite{wang2026swe,chen2025swe}. On SWE-bench Verified, we evaluate four representative model backends: GPT-5.4-mini, GPT 5.1, Gemini-3.1-Flash-Lite, and DeepSeek-v4-Flash. Due to computational costs, the analysis experiments were conducted on GPT-5.4-mini and DeepSeek-v4-Flash, while Gemini-3.1-Flash-Lite was evaluated on the 300 instances identified by our motivation experiment as having the strongest data-leakage evidence.

\subsection{Metrics}
For issue-resolution experiments on SWE-bench Verified and SWE-rebench, we use \textbf{Pass@1} as the primary effectiveness metric, defined as the proportion of benchmark instances solved by the agent in a single run. For SWE-QA, we report the benchmark's default \textbf{Average Score} metric~\cite{peng2026swe}. Across settings, we also report \textbf{Average Actions} to characterize interaction length, \textbf{Average Input Tokens} to quantify prompt-side inference cost, and \textbf{Average Output Tokens} to quantify generation-side inference cost. Together, these metrics let us evaluate not only whether \approach changes task success, but also how it affects the agent's efficiency in completing the task.


\subsection{Implementation Details}
All experiments and evaluation settings use default parameters. For \texttt{mini-swe-agent}, this means using the default configuration with decoding temperature fixed at 0 and the maximum action set to 250 per instance. 

In principle, Levels 3 and 4 could be applied to every eligible file in a repository. In practice, repository-wide transformation would introduce substantial preprocessing, validation, and runtime cost at the scale of SWE-bench Verified. We therefore restrict both levels to the files and code regions implicated by the golden patch. For Level 3, reordering is applied only within golden-patch-related files. For Level 4, functionality-preserving rewriting is applied only to golden-patch-related code regions. This design keeps the transformed view centered on the implementation most relevant to the target issue while making large-scale evaluation computationally feasible.

For each transformation level, we generate three transformed repository views for every benchmark instance using different random seeds. Agents are evaluated independently on each view, and reported results are averaged across the three runs. This protocol measures robustness across multiple semantically equivalent repository representations rather than a single fixed repository presentation. Consequently, different instances originating from the same repository are typically evaluated under different transformed repository views, and even the same instance is encountered through different repository presentations across repeated runs. This design removes dependence on a single fixed repository presentation and enables us to measure agent robustness across multiple semantically equivalent views of the same underlying task.

\section{Empirical Results}

\subsection{RQ1: Effects of \approach}

RQ1 uses SWE-bench Verified~\cite{jimenez2024swe}, a manually curated 500-instance subset of SWE-bench for repository-level issue resolution. Table~\ref{tab:rq1-main} summarizes both the joint and level-wise results. When all four transformation levels are enabled together, Pass@1 drops substantially across all evaluated models: GPT-5.4-mini decreases from 46.8\% to 35.6\% ($-11.2\%$), DeepSeek-v4-Flash from 72.8\% to 66.8\% ($-6.0\%$), GPT 5.1 from 44.6\% to 36.2\% ($-8.4\%$), and Gemini-3.1-Flash-Lite from 56.7\% to 42.3\% ($-14.4\%$). These drops are statistically significant at $p<0.05$, indicating that the performance degradation is unlikely to arise from random evaluation variation. At the same time, all models exhibit substantial increases in interaction cost, with average actions and token consumption rising markedly in the full transformed setting. Input-token usage increases by more than 2.5$\times$ for the strongest affected configurations. Because the underlying issue, execution environment, and test-defined correctness criteria remain unchanged, the observed degradation cannot be explained by changes in task semantics. Instead, it suggests that performance on canonical SWE-bench partially benefits from familiarity with canonical repository representations, and that current agents remain sensitive to repository-side cues encountered during training.


The level-wise results further reveal where this robustness gap originates. Level~1 (Problem Statement Reconstruction) has only a minor effect on Pass@1, with GPT 5.1 and GPT-5.4-mini unchanged and DeepSeek-v4-Flash decreasing by 2.0 percentage points, suggesting that simply reformulating the problem statement is insufficient to remove the familiarity advantages associated with canonical repository representations. In contrast, the largest degradation consistently comes from repository-side transformations, particularly Level~2 (Namespace Mapping), which reduces Pass@1 by 7.4, 6.4, and 6.0 percentage points for GPT 5.1, GPT-5.4-mini, and DeepSeek-v4-Flash, respectively, with statistically significant drops across all settings ($p<0.01$). Namespace Mapping also induces the largest increases in interaction cost: average actions rise by 63.3\%, 32.6\%, and 112.4\%, while input-token usage rises by 162.3\%, 108.2\%, and 218.3\% for the same three models. Levels~3 and~4 exhibit milder effects, with Pass@1 changes ranging from 0.8 to 3.4 percentage points for Level~3 and from 1.2 to 2.8 percentage points for Level~4, suggesting that modifications to local file organization and implementation patterns are less disruptive than changes to repository-level naming structure, although they still increase interaction cost. Nevertheless, these transformations remain important because they remove additional repository-side cues beyond those affected by Level~2. The full \approach setting produces the largest overall degradation, indicating that familiarity with canonical repository representations is distributed across multiple levels of repository presentation rather than concentrated in naming structure alone. Taken together, these results suggest that the observed degradation is driven primarily by changes in repository representation rather than changes to the underlying executable task itself.

Overall, these findings indicate that repository representation plays a central role in repository-level coding-agent evaluation. Performance under a single canonical repository presentation may therefore conflate representation-robust repository reasoning with sensitivity to specific repository-side representations.

\begin{findingbox}{Answer to RQ1}
Removing familiarity with canonical repository representations causes statistically significant performance degradation across LLMs (drop by 6.0\%--14.4\%, $p < 0.05$), suggesting that strong SWE-bench performance partially relies on repository-side cues beyond repository-level reasoning alone.
\end{findingbox}


\subsection{RQ2: Effect on Agent Behavior During Issue Resolution}

\begin{figure}[t]
\centering
\includegraphics[width=\columnwidth]{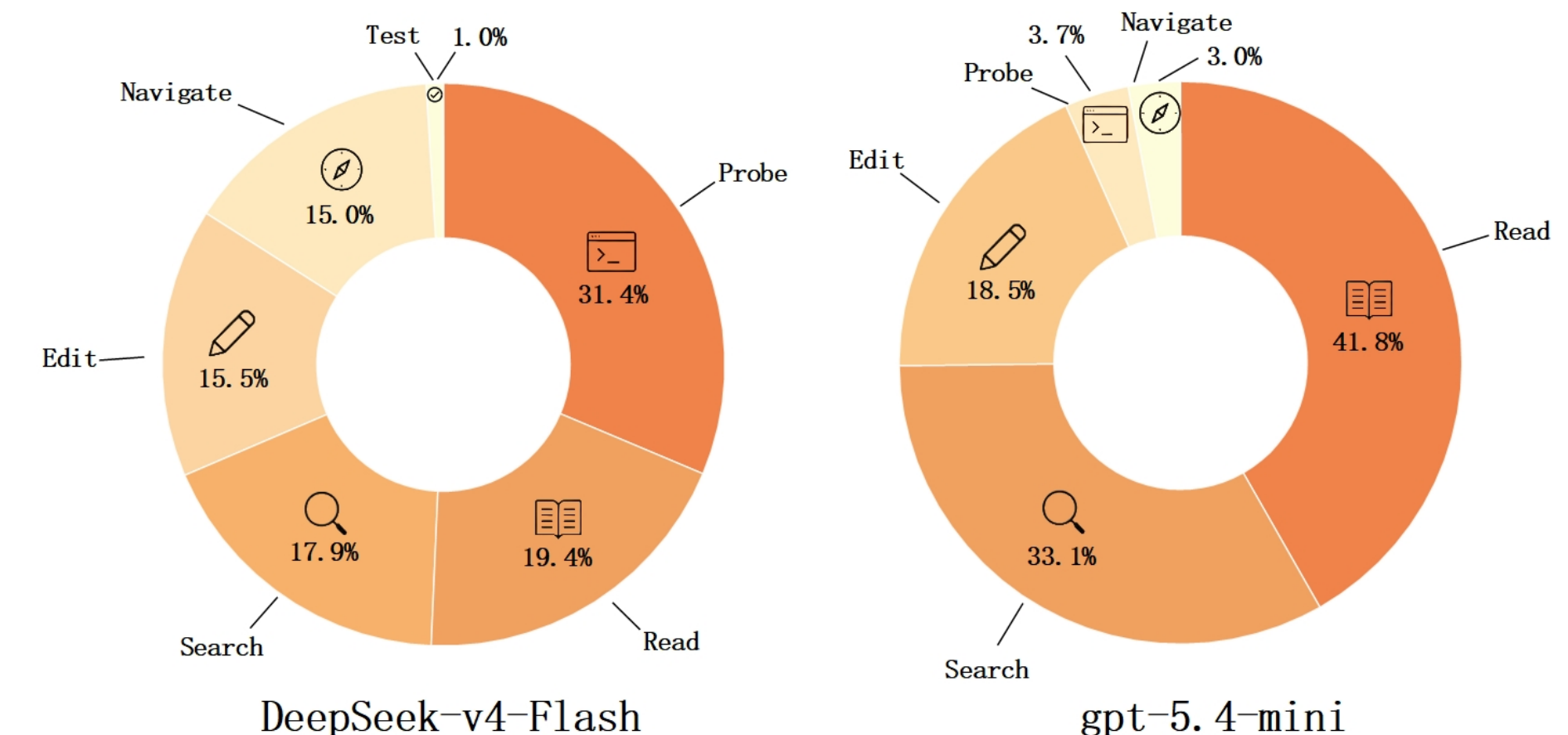}
\caption{Actions increase over baseline across categories.
}
\label{fig:rq2-action-shift}
\vspace{-8pt}
\end{figure}

RQ2 analyzes agent trajectories on SWE-bench Verified~\cite{jimenez2024swe}, using the same issue-resolution setting as RQ1. Figure~\ref{fig:rq2-action-shift} focuses on the full \approach setting and shows how the additional actions over the clean baseline are redistributed across the six action categories. For both models, the extra interaction budget is dominated by localization- and exploration-related actions rather than downstream modification or validation. For DeepSeek-v4-Flash, 83.6\% of the additional actions fall into \texttt{navigate}, \texttt{search}, \texttt{read}, and \texttt{probe}, while only 16.5\% are allocated to \texttt{edit} and \texttt{test}. For GPT-5.4-mini, a highly similar pattern emerges, with 81.6\% of the additional actions again concentrated in \texttt{navigate}, \texttt{search}, \texttt{read}, and \texttt{probe}. This consistency across models aligns with the interpretation from RQ1: when repository-side cues are eroded, agents cannot quickly localize the fault and directly determine the required fix, and must instead expend additional effort on reconstructing contextual grounding through exploration.

Despite this shared high-level shift, the two models exhibit distinct exploration strategies. DeepSeek-v4-Flash allocates a larger proportion of additional actions to \texttt{probe} (31.4\%), followed by \texttt{read} (19.4\%) and \texttt{search} (17.9\%), indicating a preference for lightweight runtime execution to actively interrogate repository behavior. In contrast, GPT-5.4-mini concentrates its exploration on \texttt{read} (41.8\%) and \texttt{search} (33.1\%), while \texttt{probe} remains minimal (3.7\%), suggesting a more conservative strategy that relies primarily on static inspection and textual retrieval rather than executable probing. GPT-5.1 exhibits an action distribution highly similar to GPT-5.4-mini, particularly in its dominance of \texttt{read} and \texttt{search} over \texttt{probe}. For brevity, its results are not separately visualized in Figure~\ref{fig:rq2-action-shift}.

The models also respond differently to transformed repository views. DeepSeek-v4-Flash adopts a more inherently exploratory interaction strategy, already issuing substantially more actions in the baseline setting and further increasing exploration under transformed repository views. This aggressive exploration behavior helps preserve a larger fraction of its original Pass@1 performance, albeit at the cost of significantly higher interaction overhead and token consumption. In contrast, GPT-5.4-mini follows a more interaction-efficient strategy, with fewer exploratory actions even under transformed views, but consequently experiences a larger relative performance degradation once familiar repository-side cues are eroded. This indicates a trade-off between exploration intensity and robustness: more exploratory agents are better able to recover missing repository context, while more conservative agents are more sensitive to representation shifts.

Taken together, the results indicate that the primary behavioral cost of \approach is not increased editing or testing effort, but a systematic reallocation of actions toward repository exploration. This shift exposes a fundamental coupling between repository representation quality and exploration policy: when familiar cues are removed, agents must rebuild repository understanding through navigation, search, reading, and probing rather than directly executing fixes.

\begin{table}[t]
  \centering
  \caption{Results on SWE-QA. \approach lowers score and increases actions and token cost.}
  \label{tab:rq3-sweqa}
  \renewcommand{\arraystretch}{1.08}
  {\scriptsize
  \setlength{\tabcolsep}{2.5pt}
  \resizebox{\columnwidth}{!}{%
  \begin{tabular}{l c c c c}
  \hline
  \multicolumn{5}{c}{\textbf{GPT-5.4-mini}} \\
  \hline
  Setting & Score & Actions & Input Tokens & Output Tokens \\
  \hline
  Baseline & 70.35 & 6.28 & 30,129.88 & 891.28 \\
  \hline
  Level 1 & \cellchange{70.32}{\gooddown{0.03}} & \cellchange{6.25}{\baddown{0.48\%}} & \cellchange{30,012.37}{\baddown{0.39\%}} & \cellchange{897.44}{\goodup{0.69\%}} \\
  \hline
  Level 2 & \cellchange{66.40}{\gooddown{3.95}} & \cellchange{7.41}{\goodup{18.00\%}} & \cellchange{38,716.90}{\goodup{28.50\%}} & \cellchange{1,106.97}{\goodup{24.20\%}} \\
  \hline
  Level 3 & \cellchange{69.30}{\gooddown{1.05}} & \cellchange{7.17}{\goodup{14.20\%}} & \cellchange{35,734.04}{\goodup{18.60\%}} & \cellchange{1,037.45}{\goodup{16.40\%}} \\
  \hline
  Level 4 & \cellchange{69.50}{\gooddown{0.85}} & \cellchange{6.96}{\goodup{10.80\%}} & \cellchange{33,956.37}{\goodup{12.70\%}} & \cellchange{1,009.82}{\goodup{13.30\%}} \\
  \hline
  \textbf{\approach} & \cellchange{65.71}{\gooddown{4.64}} & \cellchange{7.42}{\goodup{18.15\%}} & \cellchange{40,567.00}{\goodup{34.65\%}} & \cellchange{1,081.31}{\goodup{21.32\%}} \\
  \hline
  \multicolumn{5}{c}{\textbf{DeepSeek-v4-Flash}} \\
  \hline
  Setting & Score & Actions & Input Tokens & Output Tokens \\
  \hline
  Baseline & 72.97 & 24.49 & 313,610.13 & 4,369.05 \\
  \hline
  Level 1 & \cellchange{73.05}{\badup{0.08}} & \cellchange{24.31}{\baddown{0.73\%}} & \cellchange{315,803.40}{\goodup{0.70\%}} & \cellchange{4,342.84}{\baddown{0.60\%}} \\
  \hline
  Level 2 & \cellchange{72.52}{\gooddown{0.45}} & \cellchange{34.58}{\goodup{41.20\%}} & \cellchange{490,737.13}{\goodup{56.48\%}} & \cellchange{5,711.22}{\goodup{30.72\%}} \\
  \hline
  Level 3 & \cellchange{72.91}{\gooddown{0.06}} & \cellchange{27.02}{\goodup{10.33\%}} & \cellchange{361,341.59}{\goodup{15.22\%}} & \cellchange{5,013.92}{\goodup{14.76\%}} \\
  \hline
  Level 4 & \cellchange{73.01}{\badup{0.04}} & \cellchange{24.55}{\goodup{0.24\%}} & \cellchange{314,927.29}{\goodup{0.42\%}} & \cellchange{4,468.66}{\goodup{2.28\%}} \\
  \hline
  \textbf{\approach} & \cellchange{72.42}{\gooddown{0.75}} & \cellchange{35.02}{\goodup{43.02\%}} & \cellchange{498,951.10}{\goodup{59.10\%}} & \cellchange{5,770.18}{\goodup{32.07\%}} \\
  \hline
  \end{tabular}
  }
  }
\end{table}

\begin{findingbox}{Answer to RQ2}
More of the extra actions (more than 80\%) over baseline are spent on repository exploration for both models when LLMs are not familiar with the clues on the surface of the repository.
\end{findingbox}

\subsection{RQ3: Transfer to Other Tasks}

RQ3 uses SWE-QA~\cite{peng2026swe}, a repository-level question-answering benchmark. We evaluate three repositories, \texttt{Conan}, \texttt{Reflex}, and \texttt{Streamlink}, with 48 instances per repository and 144 instances in total.

Table~\ref{tab:rq3-sweqa} shows that the effects of \approach extend beyond issue resolution to repository-level question answering. On SWE-QA, GPT-5.4-mini drops from 70.35 to 65.71 in average score, while DeepSeek-v4-Flash shows only a marginal decrease from 72.97 to 72.42. Similar to the issue-resolution results in RQ1, the largest impact again comes from Level~2 (Namespace Mapping), whereas Level~1 (Problem Statement Reconstruction) produces almost no effect and Levels~3 and~4 lead to only minor changes in answer quality. This pattern suggests that repository-owned names remain the dominant repository-side cues even for question-answering tasks.

At the same time, both models exhibit substantially increased interaction cost under alternative repository representations. For GPT-5.4-mini, average actions increase from 6.28 to 7.42, accompanied by higher token consumption. For DeepSeek-v4-Flash, actions increase from 24.49 to 35.02, with a 59.10\% increase in input tokens. Consistent with RQ2, alternative repository representations primarily increase the effort required to recover repository context rather than the effort spent on producing final answers.

Compared with repository-level issue resolution, however, the effects of Levels~3 and~4 are noticeably smaller. One reason is that repository question answering mainly requires locating and understanding relevant repository context, without the full workflow of fault localization, code modification, test execution, and iterative validation. Consequently, changes to local implementation structure and functionality-preserving rewrites have less opportunity to influence agent behavior once the relevant repository context has been identified.

The two models again exhibit distinct robustness profiles consistent with RQ2. Models with more exploratory interaction strategies, such as DeepSeek-v4-Flash, are able to reconstruct sufficient repository context under alternative views, thereby largely preserving QA performance at higher interaction cost. In contrast, more efficient agents such as GPT-5.4-mini construct working context more efficiently under the canonical repository representation but become more sensitive to alternative instantiations, leading to both increased interaction cost and degraded answer quality.

Overall, these results indicate that the behavioral shift identified in RQ2 generalizes to other repository-level tasks. When canonical repository representations are replaced by alternative instantiations, agents must spend additional effort reconstructing repository context, demonstrating that reliance on familiar repository-side cues extends beyond issue-resolution tasks.

\begin{figure*}[t]
  \centering
  \includegraphics[width=0.99\textwidth]{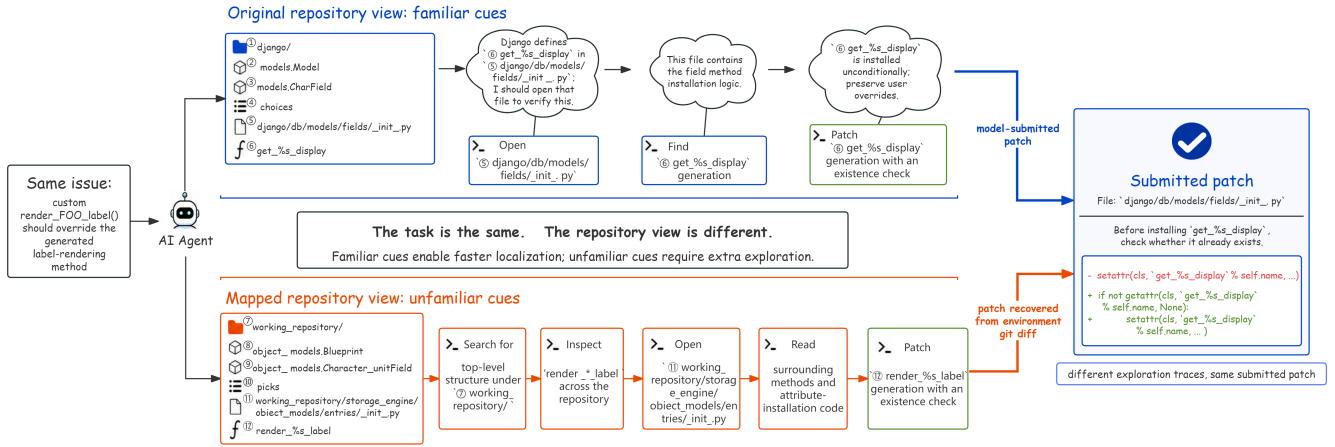}
  \caption{Case study of different agent trajectories under familiar and unfamiliar repository views.}
  \label{fig:case-study}
\end{figure*}

\begin{findingbox}{Answer to RQ3}
The behavioral and efficiency shifts induced by \approach consistently generalize to repository-level question answering, demonstrating its impact on other repository-level tasks.
\end{findingbox}

\begin{table}[t]
  \centering
  \caption{Results on the March 2026 SWE-rebench Leaderboard.}
  \label{tab:rq4-swerebench}
  \renewcommand{\arraystretch}{1.08}
  {\scriptsize
  \setlength{\tabcolsep}{2.5pt}
  \resizebox{\columnwidth}{!}{%
  \begin{tabular}{l c c c c}
  \hline
  \multicolumn{5}{c}{\textbf{GPT-5.4-mini}} \\
  \hline
  Setting & Pass@1 & Actions & Input Tokens & Output Tokens \\
  \hline
  Baseline & 17.27\% & 15.82 & 143,626.49 & 2,842.78 \\
  \textbf{\approach} & \cellchange{17.27\%}{0.00\%} & \cellchange{17.11}{\goodup{8.15\%}} & \cellchange{175,231.61}{\goodup{22.01\%}} & \cellchange{3,285.61}{\goodup{15.58\%}} \\
  \hline
  \end{tabular}
  }
  }
  \end{table}

\subsection{RQ4: Does \approach Increase Task Difficulty?}
RQ4 uses the March 2026 SWE-rebench Leaderboard split~\cite{badertdinov2026swe}, which contains 110 instances created after the release of GPT-5.4-mini. Table~\ref{tab:rq4-swerebench} shows a different pattern from SWE-bench Verified. GPT-5.4-mini maintains the same Pass@1 of 17.27\% under the full \approach setting, despite noticeable increases in interaction cost. Specifically, average actions increase from 15.82 to 17.11, input tokens increase by 22.01\%, and output tokens increase by 15.58\%.
The unchanged pass rate indicates that \approach does not make the underlying issue-resolution tasks more difficult in terms of solvability. Instead, it changes how agents interact with the repository during problem solving, leading to a higher exploration cost while preserving final outcomes.

We further observe that the increase in interaction cost is more pronounced in repositories that are widely used and long-standing, which are more likely to have been encountered during model pretraining or development. This further suggests that canonical SWE-bench instances may partially benefit from agents’ prior familiarity with repository-specific surface representations, such as naming conventions and structural patterns. By introducing alternative repository view instantiations, \approach reduces reliance on such familiarity and elicits interaction behaviors that better reflect repository-level reasoning under previously unseen representations.

Importantly, these results indicate that the observed performance differences are not due to increased task difficulty, but rather stem from changes in the required interaction process under alternative repository views.

\begin{findingbox}{Answer to RQ4}
\approach primarily removes reliance on familiar repository-side cues rather than increasing task difficulty. On temporally held-out instances, Pass@1 remains 17.27\%, while actions and input tokens increase by 8.15\% and 22.01\%, respectively.
\end{findingbox}

\subsection{Case Study: Familiar vs. Unfamiliar Repository Views}

Figure~\ref{fig:case-study} illustrates how repository-side surface cues affect agent behavior on \texttt{django\_\_django-11999}. The issue concerns a user-defined \texttt{get\_FOO\_display()} method being overwritten by Django's generated display method. In the original repository view, the agent quickly follows the familiar Django terminology\textsuperscript{1,2,3,4} in the issue: after listing the repository, it searches for \texttt{get\_.*\_display}\textsuperscript{6} under \texttt{django/db/models/}, which immediately returns \texttt{django/db/models/fields/\_\_init\_\_.py}\textsuperscript{5} and \texttt{django/db/models/base.py}. It then opens the field file around the matched lines and observes that \texttt{get\_\%s\_display}\textsuperscript{6} is installed unconditionally with \texttt{setattr}. After briefly checking the backing implementation in \texttt{base.py}, the agent has already localized the faulty code and identifies the required fix: guard the generated method installation with an existence check.

Under \approach, the same issue is exposed through the transformed repository view. The agent can no longer rely on the original Django path and class names\textsuperscript{8,9}, so its trajectory begins with broader repository exploration: it lists the current directory, enumerates Python files, inspects the top-level \texttt{working\_repository/}\textsuperscript{7} tree, and searches both source and test files for display-related methods. Only after this exploration does it open the transformed implementation file, \texttt{working\_repository/storage\_engine/\allowbreak object\_models/entries/\_\_init\_\_.py}\textsuperscript{11}, along with the related backing-method file \texttt{working\_repository/storage\_engine/\allowbreak object\_models/anchor.py}. Even after finding the analogous \texttt{render\_\%s\_label}\textsuperscript{12} generation logic, the agent continues to inspect surrounding field-registration logic\textsuperscript{10}, tests, configuration files, and a minimal reproduction setup to confirm that the transformed names correspond to the original display-method behavior. Consequently, the transformed run requires substantially more interaction steps than the original run (\(37 \rightarrow 217\) actions), while still converging to the same submitted patch. This case concretely illustrates the mechanism behind our quantitative findings: \approach changes the interaction process by removing familiar repository-side representations, rather than making the underlying repair intrinsically different.


\section{Threats to Validity} 

\paragraph{External Validity}
Our evaluation is conducted on SWE-bench Verified and SWE-QA, which are primarily composed of Python-based, open-source repositories. While these benchmarks provide realistic repository-level software engineering tasks, they may still reflect ecosystem-specific characteristics and SWE-bench-style interaction patterns. As a result, it remains an open question whether the observed robustness gaps and behavioral shifts generalize to other programming languages and repository ecosystems with different structural conventions and development practices.
 
\paragraph{Internal Validity}
Our current implementation primarily targets command-line-based repository interaction. While this setup covers the majority of operations in the evaluated agents, agents equipped with richer repository-aware tools (e.g., IDE APIs, language-server queries, or AST-level navigation) may require additional adaptation to ensure that transformed repository representations remain consistent across all tool interfaces. Extending \approach to support such tools is an important direction for future work.

\section{Related Work}

\subsection{SWE-bench Variants}

Repository-level software engineering evaluation has expanded rapidly around SWE-bench, which turns real GitHub issues into executable patch-generation tasks\cite{jimenez2024swe,oliva2025spice,fan2025swe,hu2026repo2run,jain2025r2e,wang2025swe,shi2026swebenchpromaxbenchmarkingagents}. SWE-bench Verified improves reliability through manual validation, making it a standard benchmark for frontier coding agents\cite{jimenez2024swe}. SWE-Bench+ further enhances SWE-style evaluation by improving benchmark instance quality and metadata coverage\cite{aleithan2024swe}. SWE-bench Live and SWE-rebench emphasize temporal freshness by incorporating newly released or post-release tasks to reduce direct contamination\cite{zhang2026swe,badertdinov2026swe}. SWE-Bench Pro shifts toward more complex, long-horizon issues from actively maintained repositories, while SWE-bench Multimodal extends evaluation to settings involving mixed textual and visual inputs\cite{deng2025swe,yang2024swebenchmultimodal}.

Other work improves the construction and use of SWE-style benchmarks. SPICE automatically labels SWE-bench instances along issue clarity, test coverage, and effort dimensions, while SWE-Effi re-evaluates agent effectiveness under resource constraints\cite{oliva2025spice,fan2025swe}. Repo2Run and R2E-Gym focus on scalable executable environments and procedural training/evaluation settings, and SWE-Bench++ studies scalable generation of repository-level benchmark instances from open-source projects\cite{hu2026repo2run,jain2025r2e,wang2025swe}.

Despite these extensions, existing SWE-bench variants face a trade-off between coverage and recency. Benchmarks that prioritize newly introduced or manually curated instances improve evaluation cleanliness, but often reduce the diversity of repository-level error patterns by discarding long-tail cases present in large, mature repositories. As a result, no single variant fully captures the breadth of real-world software engineering failure modes.

This limitation motivates our approach, which preserves existing benchmark coverage while systematically varying repository presentation at evaluation time, enabling robustness assessment without sacrificing instance diversity.

\subsection{Static Benchmark Perturbation}

Recent work has investigated whether strong performance on SWE-bench reflects genuine repository-level reasoning or sensitivity to fixed benchmark artifacts\cite{wang2025solved,ahmed2026investigating,yu2025utboost,yu2026swe,garg2025saving,ma2026same,10.1007/s10664-026-10802-w}. SWE-Bench Illusion shows that models can exploit stable associations between issue descriptions and repository locations, suggesting that performance may partially rely on learned correlations in static evaluation settings\cite{liang2025swe}. LastingBench studies benchmark construction strategies for defending against knowledge leakage, reinforcing the need to make benchmark instances less reusable as memorized artifacts\cite{fang2025lastingbench}. 

Several studies examine whether test-based success on SWE-bench faithfully captures correct issue resolution. Wang et al. revisit solved SWE-bench issues and question whether accepted patches are always semantically correct, while Ahmed et al. analyze test overfitting as a source of inflated benchmark performance\cite{wang2025solved,ahmed2026investigating}. UTBoost and SWE-ABS strengthen evaluation through more rigorous or adversarial tests, and Saving SWE-Bench mutates benchmark instances to create more realistic agent evaluation settings\cite{yu2025utboost,yu2026swe,garg2025saving}.

PoorCodeSumEval introduces controlled obfuscations of code by modifying identifiers and reducing readability while preserving program behavior, aiming to test robustness to superficial lexical cues\cite{10.1145/3691620.3695072}. RepoMirage applies repository-level perturbations to SWE-bench instances to increase apparent difficulty under altered structural presentations\cite{li2026repomirage}. 

Together, these approaches demonstrate that modifying static benchmark presentations can expose sensitivity to surface-level regularities in evaluation artifacts. However, because the transformed instances are fixed once constructed, they may still be incorporated into future training or evaluation corpora, limiting their ability to prevent repeated exposure to specific representations.

In contrast, SchrodingerRepo instantiates repository representations as latent, evaluation-time variables, generating fresh, seed-conditioned repository views instantiated for each evaluation run while preserving full execution semantics and correctness criteria. This design ensures that agents are evaluated on whether they can genuinely resolve the underlying issue from the realized repository structure, rather than relying on prior familiarity with canonical repository presentations.

\section{Conclusion}

Repository-level benchmarks such as SWE-bench expose each task through a single fixed repository presentation, making it difficult to separate robust repository reasoning from familiarity. We introduced \approach, a semantics-preserving framework that instantiates repository representations only at evaluation time. On SWE-bench Verified, full transformations lower Pass@1 by 6.0--14.4 percentage points across models. Trajectory analysis shows that 81.6--83.6\% of the extra actions are spent on exploration and localization, rather than editing or testing. The effect also transfers to SWE-QA, where scores drop by 0.75--4.64 points and actions increase by 18.15--43.02\%, indicating that current coding agents rely substantially on familiar repository-side cues. These results demonstrate the importance of treating repository representation as an explicit experimental variable when evaluating coding agents. This shifts repository-level evaluation from measuring performance on a single potentially familiar presentation toward measuring robustness across semantically equivalent repository realizations. Future work is to support richer repository-aware tool interfaces, including IDE APIs and language-server queries, so that representation-robust evaluation remains consistent beyond command-line agent workflows.



\bibliographystyle{IEEEtran}
\bibliography{ref}

\end{document}